\documentclass[
  ,draft            
  ]
  {aipproc}
\layoutstyle{6x9}
\usepackage{graphicx,amssymb}

\begin{document}

\title{Thermalization and Flow of Heavy Quarks in the Quark-Gluon
  Plasma}


\author{Hendrik van Hees}{address={Cyclotron Institute and Physics
    Department, Texas A{\&}M University, College Station, Texas
    77843-3366, USA}}

\author{Vincenzo Greco}{address={Laboratori Nazionali del Sud INFN, via
    S. Sofia 62, I-95123 Catania, Italy}}

\author{Ralf Rapp}{address={Cyclotron Institute and Physics
    Department, Texas A{\&}M University, College Station, Texas
    77843-3366, USA}}

\date{January 19, 2006}


\begin{abstract}
  Elastic scattering of charm ($c$) and bottom ($b$) quarks via $D$- and
  $B$-meson resonance states in an expanding, strongly interacting
  quark-gluon plasma is investigated. Drag and diffusion coefficients
  are calculated from an effective model based on chiral symmetry and
  heavy-quark effective theory, and utilized in a relativistic Langevin
  simulation to obtain transverse-momentum spectra and elliptic flow
  ($v_2$) of $c$- and $b$-quarks. The hadronization to $D$- and
  $B$-mesons is described by coalescence and fragmentation, and the
  resulting decay-electron spectra are compared to recent RHIC data.
\end{abstract}

\maketitle

\textbf{Introduction.} Recent experimental results at the Relativistic
Heavy-Ion Collider (RHIC) have given convincing evidence for the
creation of dense partonic matter with large collectivity and opacity. A
key challenge in the description of this strongly interacting
quark-gluon plasma (sQGP) is the understanding of the microscopic
reaction mechanisms, leading to its approximate behavior as a nearly
perfect fluid.
 
Heavy quarks (HQs) are valuable probes for the properties of the dense
matter produced in heavy-ion reactions, since they are expected to be
created in the early stages of the collision. Recent measurements of the
transverse-momentum ($p_T$) spectra of non-photonic single electrons
($e^{\pm}$) at RHIC, attributed to the decay of $D$- and $B$-mesons,
show a surprisingly small nuclear modification factor,
$R_{AA}^{e}$~\cite{phenix-e1,jac05,bil05}, and large elliptic flow,
$v_2^{e}$~\cite{v2-phenix,v2pre-star,aki05}.  To explain these findings,
especially the large $v_2^{(e)}$, quark-coalescence
models~\cite{GKR04,mol05,ZCK05} require that charm quarks are in
approximate thermal equilibrium with light partons. A large degree of
$c$-quark thermalization is, however, not supported by approaches based
on perturbative Quantum Chromodynamics (pQCD), e.g., using radiative
energy-loss~\cite{DGW05,arm05}. While at lower $p_T$ elastic scattering
processes parametrically dominate the energy loss ($\sim
1/\sqrt{\alpha_s}$)~\cite{MT05}, a $c$-quark $R_{AA}$ compatible with
the observed $R_{AA}^{e}$ can only be obtained with unrealistically
large values of the strong coupling constant~\cite{MT05}. Also the
combined effects of elastic and radiative energy loss may not explain
the experimental findings~\cite{WHDG05}.

In this talk we introduce $D$- and $B$-meson like resonance states in
the sQGP~\cite{HR04} mediating elastic rescattering for heavy quarks.
Employing pertinent drag and diffusion coefficients within a
Fokker-Planck approach~\cite{sve88}, we calculate HQ distributions in a
flowing thermal QGP to simulate semi-central Au-Au collisions at
RHIC~\cite{HGR05}. Hadronization to $D$- and $B$-mesons is described by
a combined quark-coalescence and fragmentation model, and subsequent
semileptonic decay electron spectra are compared to recent data.

\textbf{Heavy-quark rescattering in the QGP.} Lattice QCD (lQCD)
computations of hadronic correlators and lQCD-based effective models
suggest that mesonic resonance/bound states survive in the QGP up to
temperatures of $\sim 2 T_c$ in the light- and heavy-quark
sector~\cite{AH04}. We here assume that the lowest pseudoscalar $D$- and
$B$-meson states persist above the heavy-light quark
threshold~\cite{HR04}. Chiral and HQ symmetry imply the degeneracy with
scalar, vector and axial-vector states. Pertinent resonant $Q$-$\bar{q}$
cross sections are supplemented with leading-order pQCD
processes~\cite{com79}, using $\alpha_s=g^2/(4 \pi)=0.4$. The evaluation
of drag and diffusion coefficients within a Fokker-Planck
model~\cite{sve88} results in HQ thermalization times which are lower by
a factor $\sim 3$ compared to pQCD scattering~\cite{HGR05}.

These coefficients are used in a relativistic Langevin
simulation~\cite{MT05} for the rescattering of HQs in an isentropically
expanding QGP fireball corresponding to $b=7$~fm Au-Au collisions at
RHIC. The expansion parameters are determined to resemble the time
evolution of radial and elliptic flow in hydrodynamic
models~\cite{KSH00}, with an ideal QGP equation of state with $2.5$
flavors and a formation time of $1/3$~fm/c (initial temperature
$T_0=340$~MeV). The proper thermal equilibrium limit in the Langevin
process is implemented via the H{\"a}nggi-Klimontovich
realization~\cite{DH05}, with longitudinal diffusion coefficient $B_1=T
E A$~\cite{MT05} (Einstein's dissipation-fluctuation relation).

The initial HQ-$p_T$-distributions and the relative magnitude of $c$-
and $b$-quark spectra are determined by fitting experimental $D$ and
$D^*$ spectra in d-Au collisions~\cite{dAu05}.  The corresponding
$e^{\pm}$ spectra saturate data from $p$-$p$ and d-Au for $p_T^{e}
\lesssim 3.5$~GeV~\cite{dAu05,ppel05} with the missing yield at
higher $p_T$ attributed to $B$-meson decays, leading to a cross-section
ratio of $\sigma_{b\bar{b}}/\sigma_{c\bar{c}} = 4.9 \cdot 10^{-3}$ and a
crossing of $D$- and $B$-decay electrons at $p_T \simeq 5$~GeV.
\begin{figure}
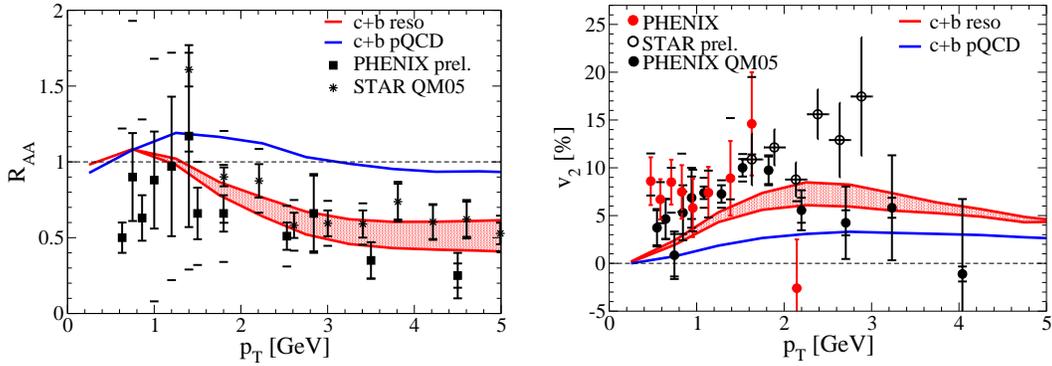

\begin{minipage}{0.45 \textwidth}
\includegraphics[width=\textwidth]{raa_e_minbias_wfrag-therm-av}
\end{minipage}\hspace*{5mm}
\begin{minipage}{0.45 \textwidth}
\includegraphics[width=\textwidth]{v2_e_minbias_wfrag-therm-av}
\end{minipage}
\caption{Nuclear modification factor, $R_{AA}$ (left panel), and
  elliptic flow, $v_2$ (right panel), of semileptonic $D$- and $B$-meson
  decay electrons in $b$$=$7~fm, $\sqrt{s_{NN}}=200$~GeV Au-Au
  collisions assuming different elastic HQ interactions in the QGP with
  subsequent coalescence, including the thermal weight factor described
  in the text, and fragmentation hadronization, compared to PHENIX and
  STAR data~\cite{phenix-e1,jac05,v2pre-star,aki05}.\label{fig.1}}
\end{figure}

\textbf{Hadronization and single-electron observables.} The $c$- and
$b$-quark spectra from the Langevin simulation are used in the
coalescence model of Ref.~\cite{GKR04} with light-quark distributions
from~\cite{GKL03}. Here we take into account the thermal weight factor
for the production of $D^*$ mesons relative to $D$ mesons,
$(m_{D^*}/m_D)^{3/2} \exp[-(m_{D^*}-m_D)/T]$, as described
in~\cite{GKR04}. This leads to a reduced fraction of $c$-quarks which
hadronize to $D^*$ mesons via coalescence, compared to our analysis
in~\cite{HGR05}. To conserve $c$- and $b$-number unpaired HQs are
hadronized via $\delta$-function fragmentation. Finally, the
single-$e^{\pm}$ are obtained from $D$- and $B$-meson three-body decays.
Fig.~\ref{fig.1} shows that resonance scattering leads to a substantial
increase in $v_2^e$ and decrease in $R_{AA}^e$, as compared to pQCD
rescattering alone.  Coalescence further amplifies $v_2^e$ but also
increases $R_{AA}^e$. The $B$-meson contributions reduce the effects for
$p_T \gtrsim 3$~GeV.

Note that the nonperturbative resonance formation mechanism employed in
this work importantly resides on a finite (equilibrium) abundance of
(anti-) quarks, while perturbative calculations~\cite{arm05,DGW05}
typically assume a maximum of color charges entirely residing in gluons.

\textbf{Conclusions.} Assuming the survival of $D$- and $B$-meson
resonances in the sQGP, we have evaluated $c$- and $b$-quark spectra in
an expanding fireball at RHIC within a relativistic Langevin simulation.
The elastic resonance rescattering of $c$-quarks leads to an $R_{AA}$
down to 0.2 and $v_2$ up to $10\%$, while $b$-quarks are less affected.
The HQs were hadronized in a combined quark-coalescence and
fragmentation model followed by semileptonic $D$- and $B$-meson decay.
The resulting $R_{AA}^{e}$ and $v_2^{e}$ are in reasonable agreement
with recent RHIC data, suggesting that HQ-interactions via resonances
may play an important role in the understanding of the microscopic
properties of the sQGP, especially the rapid thermalization of heavy
quarks.

\textbf{Acknowledgments.} One of us (HvH) has been supported in part by
a F.-Lynen Fellowship of the A.-v.-Humboldt Foundation. This work has
been supported in part by a U.S. National Science Foundation CAREER
award under grant PHY-0449489.

\begin{flushleft}

\end{flushleft}

\end{document}